\documentclass[prl,twocolumn,nofootinbib]{revtex4-1}
\usepackage{graphicx, epsfig}
\usepackage[dvipsnames]{xcolor}
\usepackage{hyperref}
\usepackage{mathrsfs}
\usepackage{amsmath}
\usepackage{enumerate}
\usepackage{bm}
\newcommand{\bmf}[1]{{\bm{#1}}}
\usepackage{float}
\usepackage{ulem}
\newcommand{\be}{\begin{equation}}
\newcommand{\ee}{\end{equation}}
\newcommand{\bea}{\begin{eqnarray}}
\newcommand{\eea}{\end{eqnarray}}
\newcommand{\ba}{\begin{eqnarray}}
\newcommand{\ea}{\end{eqnarray}}

\newcommand{\gapp}{\mathrel{\raise.3ex\hbox{$>$}\mkern-14mu
              \lower0.6ex\hbox{$\sim$}}}
\newcommand{\lapp}{\mathrel{\raise.3ex\hbox{$<$}\mkern-14mu
              \lower0.6ex\hbox{$\sim$}}}

\begin{document}
\title{Magnetic field and gravitational waves from the first-order Phase Transition}

\author{Yuefeng Di}
\affiliation{
	~Department of Physics, Chongqing University, Chongqing 401331, China}
\author{Jialong Wang}
\affiliation{
	~Department of Physics, Chongqing University, Chongqing 401331, China}
\author{Ruiyu Zhou }\affiliation{
	~Department of Physics, Chongqing University, Chongqing 401331, China}
	
\author{Ligong Bian }\email{lgbycl@cqu.edu.cn}
\affiliation{
	~Department of Physics, Chongqing University, Chongqing 401331, China}	

\author{Rong-Gen Cai}\email{cairg@itp.ac.cn}
\affiliation{CAS Key Laboratory of Theoretical Physics, Institute of Theoretical Physics, Chinese Academy of Sciences, P.O. Box 2735, Beijing 100190, China}
\affiliation{School of Physical Sciences, University of Chinese Academy of Sciences, No. 19A Yuquan Road, Beijing 100049, China}
\affiliation{School of Fundamental Physics and Mathematical Sciences, Hangzhou Institute for Advanced Study, University of Chinese Academy of Sciences, Hangzhou 310024, China}
\author{Jing Liu}\email{liujing@ucas.ac.cn}
\affiliation{School of Fundamental Physics and Mathematical Sciences, Hangzhou Institute for Advanced Study, University of Chinese Academy of Sciences, Hangzhou 310024, China}
	\affiliation{School of Physical Sciences, University of Chinese
	Academy of Sciences, Beijing 100049, China}

%%%%%%%%%%%%%%%%%%%%%%%%%%%%%%%%%%%%%%%%%%%%%%%%%%%%%%%
\begin{abstract}

We perform  the three dimensional lattice simulation of the magnetic field and gravitational wave productions from bubble collisions during the first-order electroweak phase transition. Except that of the gravitational wave, the power-law spectrum of the magnetic field strength is numerically calculated for the first time, which is of a broken power-law spectrum: $B_{\xi}\propto f^{0.91}$ for low frequency region of $f<f_\star$ and $B_{\xi}\propto f^{-1.65}$ for high frequency region of $f>f_\star$ in the thin-wall limit, with the peak frequency being $f_\star\sim 5$ Hz at the phase transition temperature 100 GeV. When the hydrodynamics is taken into account, the generated magnetic field strength can reach $B_\xi\sim 10^{-7}$G at a correlation length $\xi\sim 10^{-7}$pc, which may seed the large scale magnetic fields. Our study shows that the measurements of cosmic magnetic field strength and gravitational waves are complementary to probe new physics admitting electroweak phase transition.

\end{abstract}

\maketitle

\noindent{\it \bfseries  Introduction:}
Recent lattice simulation confirms that there is no first-order phase transition (PT) in the Standard Model (SM)~\cite{DOnofrio:2014rug}.
However, a first-order electroweak PT is a general feature in many new physics models beyond the SM, which provide one crucial Sakharov condition for the explanation of the baryon asymmetry of the Universe in the paradigm of electroweak baryogenesis\cite{Morrissey:2012db}. 
New physics models admitting a first-order PT can be explored at future high energy colliders through the probe of the Higgs pair production~\cite{Arkani-Hamed:2015vfh}.  
The study of first-order PT arises growing interest of particle physicists after the LIGO and Virgo collaborations released their direct detection of gravitational waves (GWs)~\cite{Abbott:2016blz}, since a general prediction of the first-order PT , i.e., the stochastic gravitational waves backgrounds  open a new astronomy window
to probe new physics beyond the SM~\cite{Mazumdar:2018dfl,Caprini:2015zlo,Caprini:2019egz}. The GWs from the first-order electroweak PT is an important scientific goal of Laser Interferometer Space Antenna (LISA)~\cite{Audley:2017drz}, Taiji~\cite{Guo:2018npi}, TianQin~\cite{Luo:2015ght},  Big Bang Observer (BBO)~\cite{Corbin:2005ny}, and DECi-hertz Interferometer Gravitational wave Observatory (DECIGO) ~\cite{Yagi:2011wg}. On the other hand, the ubiquitous existence of the cosmological magnetic fields (MFs) is confirmed by observations~\cite{Yamazaki:2012pg,2019MNRAS.486.4275X,2018ApJS..234...11H,doi:10.1146/annurev-astro-091916-055221}, especially the lower bounds on intergalactic MFs were derived from gamma-ray observations of blazars~\cite{Dermer:2010mm,Taylor:2011bn,Neronov:1900zz}. However, the origin of the large scale MFs is still a mystery. 
The first-order PT is proposed to produce the magnetic fields~(MFs)~\cite{Vachaspati:2001nb,Durrer:2013pga,Grasso:2000wj,Subramanian:2015lua,Kandus:2010nw,Caprini:2009yp}, which may seed the cosmic MFs. 
Vachapati~\cite{Vachaspati:2001nb} proposed to generate the MFs with the non-vanshing gradients of the Higgs fields when the bubbles collide during the first-order electroweak PT. 

In this Letter, we intend to numerically study the production of MFs and GWs during the PT process where bubble collision occurs, and investigate the possibility to probe the first-order electroweak PT directly with observations of MFs and GWs. 
We numerically calculate the MF spectrum and GW energy spectrum, where the effects of the bubble wall thickness and the Higgs gradients contribution to the MFs generation are explored. The previous study of MFs generation in Ref.\cite{Zhang:2019vsb} didn't consider the GW production. We study the generations of the MFs and GWs together for the first time. We perform a three-dimensional lattice simulation of Higgs and electroweak gauge fields rather than solely adopting a scalar theory as the previous GWs simulations~\cite{Cutting:2020nla}\footnote{Ref.~\cite{Lewicki:2020jiv,Lewicki:2020azd} studied the bubble collision in the $U(1)$ gauge theory.}. 
Our simulations consider the first-order PT admitting GWs production from bubble collisions rather than preheating in the hybrid inflation~\cite{DiazGil:2007dy,DiazGil:2008tf}. 
Since the duration time of the first-order PT is much smaller than the Hubble time, we neglect the effect of expansion of the Universe throughout this work.  

\begin{figure}[!htp]
\begin{center}
\includegraphics[width=0.45\textwidth]{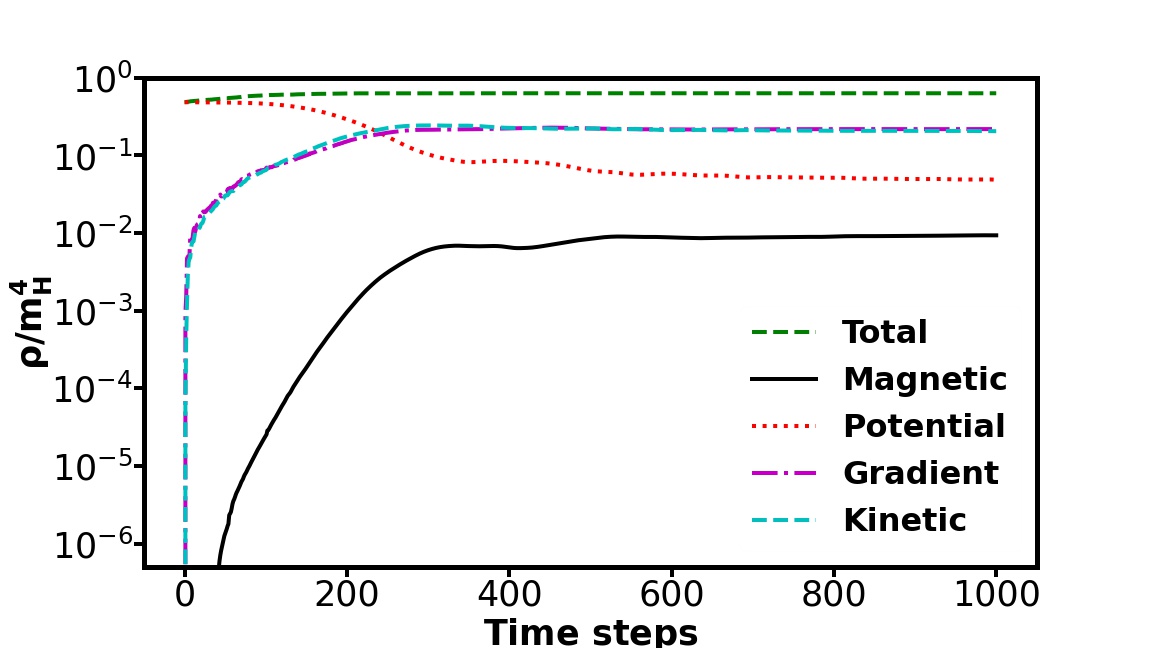}
\includegraphics[width=0.45\textwidth]{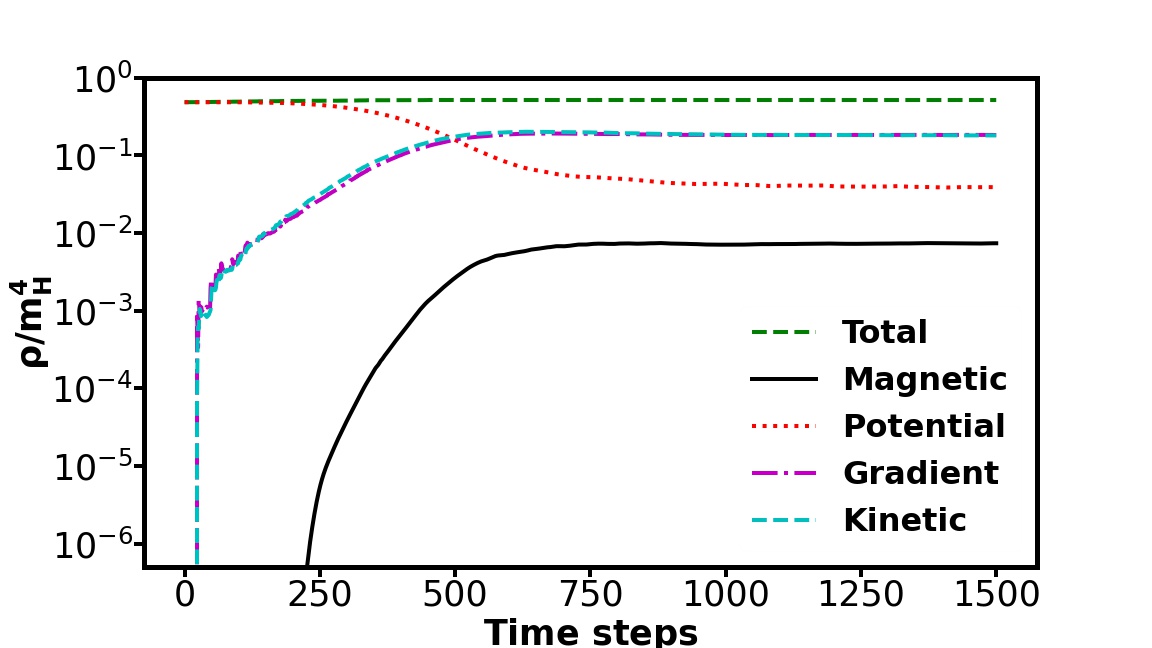}
\caption{The energy density evolution of different ingredients for bubble nucleation rates $p_B=10^{-8}$ (top panel) and $p_B=10^{-9}$ (bottom panel). } \label{fig:rhob}
\end{center}
\end{figure}

\noindent{\it \bfseries The PT model: }
\label{sec:model}
To study the generation of MFs, we firstly present the relevant Lagrangian in the electroweak theory,
\begin{equation}
\mathcal{L}=\vert D_\mu \Phi\vert^2-\frac{1}{4} W^a_{\mu\nu} W^{a\mu\nu}
                    - \frac{1}{4} B_{\mu\nu} B^{\mu\nu}-V(\Phi)\;.
\label{lagrangian}
\end{equation}
Here $\Phi$ and $V(\Phi)$ denote Higgs field and the Higgs potential, the covariant derivative is
$
D_\mu = \partial_\mu - i \frac{g}{2} \sigma^a W^a_\mu - i \frac{g'}{2} B_\mu$, where $\sigma^a$ ($a=1,2,3$) are the Pauli matrices,
and the SU(2)$_L$ and U(1)$_Y$ field strengths are $W^a_{\mu \nu}$ and $B_{\mu \nu}$, respectively.
The physical values of the coupling constants are
$g=0.65$ and $g'=0.53g$. A first-order PT can be realized in many models beyond the SM, such as: SM extended with a dimensional-six operator $(\Phi^\dagger \Phi)^3/\Lambda^2$~\cite{Grojean:2004xa,Grojean:2006bp,Cai:2017tmh},  xSM~\cite{Profumo:2014opa,Zhou:2019uzq,Zhou:2020idp,Alves:2018jsw,Profumo:2007wc,Espinosa:2011ax,Jiang:2015cwa}, 2HDM~\cite{Cline:2011mm,Dorsch:2013wja,Dorsch:2014qja,Bernon:2017jgv,Andersen:2017ika,Kainulainen:2019kyp}, George-Macheck model~\cite{Zhou:2018zli}, and NMSSM~\cite{Bian:2017wfv,Huber:2015znp}. In these models, the Higgs potential $V(\phi)$ embracing a barrier at finite temperature that can trigger a first-order PT proceeding with bubble nucleations and collisions~\cite{Grojean:2004xa} that are expected to produce MFs
~\cite{Durrer:2013pga,Kandus:2010nw,Subramanian:2015lua} and GWs~\cite{Grojean:2006bp}. The correspondence between model parameters of the $V(\phi)$ at the PT temperature and bubble characteristics (mean bubble separation, wall thickness and wall velocity) can be obtained as in Ref.~\cite{Hindmarsh:2013xza,Cutting:2019zws,Hindmarsh:2015qta,Hindmarsh:2017gnf}. 
The initial conditions are set as
$\Phi=\dot\Phi=0$, and
the profile of generated bubbles is adopted as
\begin{equation}
\label{eq:Phiini}
\Phi (t=0,{\bm r}) =  \frac{v}{2} \left [ 1-\tanh \left ( \frac{r-R_0}{L_w}\right ) \right ]
\begin{pmatrix} 0 \\ 1 \end{pmatrix} \;,
\end{equation}
where $R_0$ is the initial bubble radius and $L_w$ is  the thickness of the critical bubble wall, which is highly related with the potential barrier when the phase transition occurs.
Following Ref.~\cite{Cutting:2020nla}, we define the
``wall'' of the bubble corresponding to the section of the field
profile between $r_{In}(t)$ and $r_{Out}(t)$ where
$\phi(t,r_{In}) = v (1 - \text{tanh}\left( - 1/2 \right)) /2$ and
$\phi(t,r_{Out}) = v (1 - \text{tanh}\left( 1/2 \right)) /2$. Here
$v$ is the Higgs expectation value of the true vacuum. We consider that the  bubbles randomly nucleate in the
regions where the symmetry is unbroken to capture the dynamic of the first-order PT, where the bubble nucleation rate is directly connected with the mean bubble separation.
{\color{blue} \footnote{ See 
Supplemental Material for the relations between the parameter and particular PT models' parameter, which includes Refs.~\cite{Hindmarsh:2019phv,Cutting:2020nla,Cutting:2018tjt,Hindmarsh:2013xza}. }}
We use the temporal gauge, $W_0^a = B_0=0$, and evolve the equations of motions~(EOMs) for bosonic fields on the lattice as Ref.~\cite{Rajantie:2000nj,Zhang:2017plw} (see supplemental material) to generate the MFs and GWs. We note that in the previous simulations of GWs from first-order PT, the electroweak gauge fields are absent in the EOMs of bosonic fields, our simulation shows that its effect is indeed negligible for GWs production, due to the 
smallness of the MF energy, as can be found in Fig.~\ref{fig:rhob}. Meanwhile, the evolution of gauge fields here seeds the MFs production during the first-order PT.

\noindent{\it \bfseries MF and GW production:}
We define the electromagnetic fields after the Higgs field leaves the symmetric phase as
$
A_\mu = \sin \theta_w n^a W^a_\mu +  \cos \theta_w B_\mu$,
where $\theta_w$ is the weak mixing angle satisfying $\sin^2\theta_w=0.22$, and $n^a \equiv -(\Phi^\dagger \sigma^a \Phi)/v^2$ presents the direction of the Higgs field.
The corresponding field strength is constructed
as~\cite{tHooft:1974kcl,Vachaspati:1991nm}
\begin{equation}
\begin{split}
A_{\mu\nu} =& \sin \theta_w n^a W^a_{\mu\nu} + \cos \theta_w B_{\mu\nu} 
- i\frac{2}{g v^2} \sin \theta_w\\ &\times \left[ (D_\mu\Phi)^\dagger (D_\nu \Phi) - (D_\nu\Phi)^\dagger (D_\mu \Phi) \right]\;.
\label{eq:amunu}
\end{split}
\end{equation}
One  can see that the field strength tensor (Eq.~\ref{eq:amunu}) for the calculation of MF strength includes the contributions from the Higgs gradients, whose effects depend on the bubble collision dynamics that will be investigated latter. 
Following the conventions in Ref.~\cite{Brandenburg:2017neh, Brandenburg:2018ptt},  MFs
can be described in terms of the equal-time correlation function of $\langle B_i^* (\bmf{k},t )B_j(\bmf{k}',t)\rangle
= (2\pi )^3 \delta^{(3)}(\bmf{k}-\bmf{k}' )F_{ij} (\bmf{k},t)$,
where $ B_i (\bmf{k},t )$ is the Fourier transformation of $ B_i (\bmf{x},t )$  (the magnetic component of $A_{\mu\nu}$ in Eq.~\ref{eq:amunu}).
Since parity violation of the Lagrangian is not considered in this Letter, the antisymmetric part of $ F_{ij} (\bmf{k},t)$ vanishes, with
$
F_{ij} (\bmf{k},t)/(2\pi )^3
= (\delta_{ij}-\hat{k}_{i}\hat{k}_{j}) E_{M}(k,t)/(4\pi k^{2})$,
and the magnetic energy density is obtained as~\cite{Durrer:2013pga}:
$\rho_B \left(t\right) =\int_{0}^{\infty}E_{M}\left(k,t\right)dk$.
The MF strength can be obtained as:
$
B_\xi=\sqrt{2d\rho_B/d\log (k)}\;
$.
With the ``characteristic" correlation length being defined as,
$
\xi_M(t)=\int dk k^{-1} E_{M}\left(k,t\right)/\rho_B \left(t\right)$,
 the corresponding root mean square (scale-averaged) MF strength is~\cite{Brandenburg:2017neh} 
$
B_{rms}(t)=\sqrt{2\rho_B \left(t\right)}$.

The GWs sources from the first-order PT mainly include bubble collisions, sound waves, and turbulence. We focus on GWs produced from bubble collisions when the MFs are produced.
Recently, significant progress has been made on lattice simulations of the GW production from the first-order PT
\cite{Giblin:2014qia,Hindmarsh:2013xza,Cutting:2019zws,Hindmarsh:2015qta,Hindmarsh:2017gnf,Cutting:2018tjt,Cutting:2020nla,Pol:2019yex}.
For the calculation approach, we adopt the straightforward procedure detailed in Ref.~\cite{GarciaBellido:2007af} rather than the envelope approximation which is still under improving and cannot tell the whole story, see Ref.~\cite{Cutting:2018tjt,Cutting:2020nla} for recent lattice realization and Ref.~\cite{Konstandin:2017sat,Ellis:2020nnr,Lewicki:2020jiv,Lewicki:2020azd} for theoretical studies. 
 The EOM of tensor perturbations $h_{ij}$ reads
\begin{equation}
\ddot{h}_{ij} - \nabla^2 h_{ij} = 16 \pi G T^{\mathrm{TT}}_{ij}\;.
\end{equation}
Here the superscript $\mathrm{TT}$ denotes the transverse traceless projection, and the energy-momentum tensor is dominated by
\begin{equation}
	\begin{split}
T_{\mu\nu}=\partial_\mu \Phi^\dag \partial_\nu \Phi&-g_{\mu\nu}\frac{1}{2}\rm{Re}[(\partial_i  \Phi^\dag \partial^i \Phi )]\;,\label{ttt}
\end{split}
\end{equation}
where we neglected the contribution from the  subdominant MFs, but the MF's contribution affects the evolution of Higgs field through
 EOMs. (See supplemental material~\footnote{The GWs generated from MFs is neglegible since the energy density of MFs is subdominant. The peak of the energy spectrum of GWs from MFs is estimated as $10^{-13}$ using the method of Ref.~\cite{Caprini:2006jb}, which is much smaller than that from bubble collision, $10^{-9}$. Ref.~\cite{Saga:2018ont} presented the possibility to constraint the small scale MFs with the GWs detection with the  pulsar timing array experiments.} ).
The energy spectrum of GWs is defined as the GW energy density fraction per
logarithmic frequency interval,
\begin{equation}
	\Omega_{\mathrm{GW}}=\dfrac{1}{\rho_{c}}\frac{d\rho_\text{GW}(k)}{d\ln k}\;.
\end{equation}

\noindent{\it \bfseries Numerical results:\label{sec:Numerical-Simulation}}
Our simulations are performed on a cubic lattice with the resolution $512^{3}$, and the lattice size $L^{3}$ is related with the number of bubbles initially placed in the lattice. The time spacing is chosen to be $\Delta t=L/2048$, which is much smaller than the spacial resolution $\Delta x=L/512$. This choice of lattice spacing gives us enough resolution to ensure that we capture all the dynamics for MFs and GWs productions.
The mean bubble separation in the simulations is obtained as $R_\star=(L^3/N_b)^{1/3}$ with $N_b$ being the number of the generated bubbles at the simulation, which determines the Lorentz factor for a bubble to be $\gamma_\star=R_\star/(2 R_0)$, and the wall width as $L_w^\star=L_w/\gamma_\star$ at bubble collision time. We assume the nucleation rate is a constant during the simulation, which is denoted by $p_B$~\cite{GarciaGarcia:2016xgv}. We simulate the cases of $p_B=10^{-8}$ and $p_B=10^{-9}$ to investigate the wall thickness effects on the GWs and MFs productions. For the bubble nucleation rate $p_B=10^{-8}$, we have $\gamma_\star=2.98$, with the wall velocity being $v_w=0.94$ and the mean bubble separation $R_\star\approx 35.6 (L_w/\gamma_\star)$ at the time of bubble collision. 
To reach the thin wall limit, we consider a much lower nucleation rate $p_B=10^{-9}$, which yields: $\gamma_\star=4.84$, the wall velocity $v_w=0.98$, and $R_\star\approx 93.9 (L_w/\gamma_\star)$. 

The bubble walls are pushed outwards by the vacuum energy after bubble nucleation, and finally collide with each other.
Bubbles nucleation scenarios of
$p_B=10^{-8,-9}$ are presented to show the connection between the magnetic energy density and the bubble nucleation rate. The two panels of  Fig.~\ref{fig:rhob} show that a large $p_B$ predicts an early generation and increase of $\rho_B$, since the bubbles nucleation and collision for large $p_B$ occur early (See Fig.1 in the supplemental material for the illustration of bubble expansion and collision.). 
After the PT, $\rho_{B}$ constitutes about one percent of the total energy at the end of the simulation, which is much lower than the bounds from the Big-Bang Nucleosynthesis, $\rho_B/\rho_{rad}< 0.1$~\cite{Mukhanov:2005sc}. 
To obtain the observables at present, we red-shift the GW energy spectrum and MF spectrum by taking into account the temperature $T$ and the duration time of the PT ($\beta^{-1}$), which depend on the underlying PT models, and we set them to be free parameters in this work. 
 
  \begin{figure}[!ht]
\begin{center}
\includegraphics[width=0.4\textwidth]{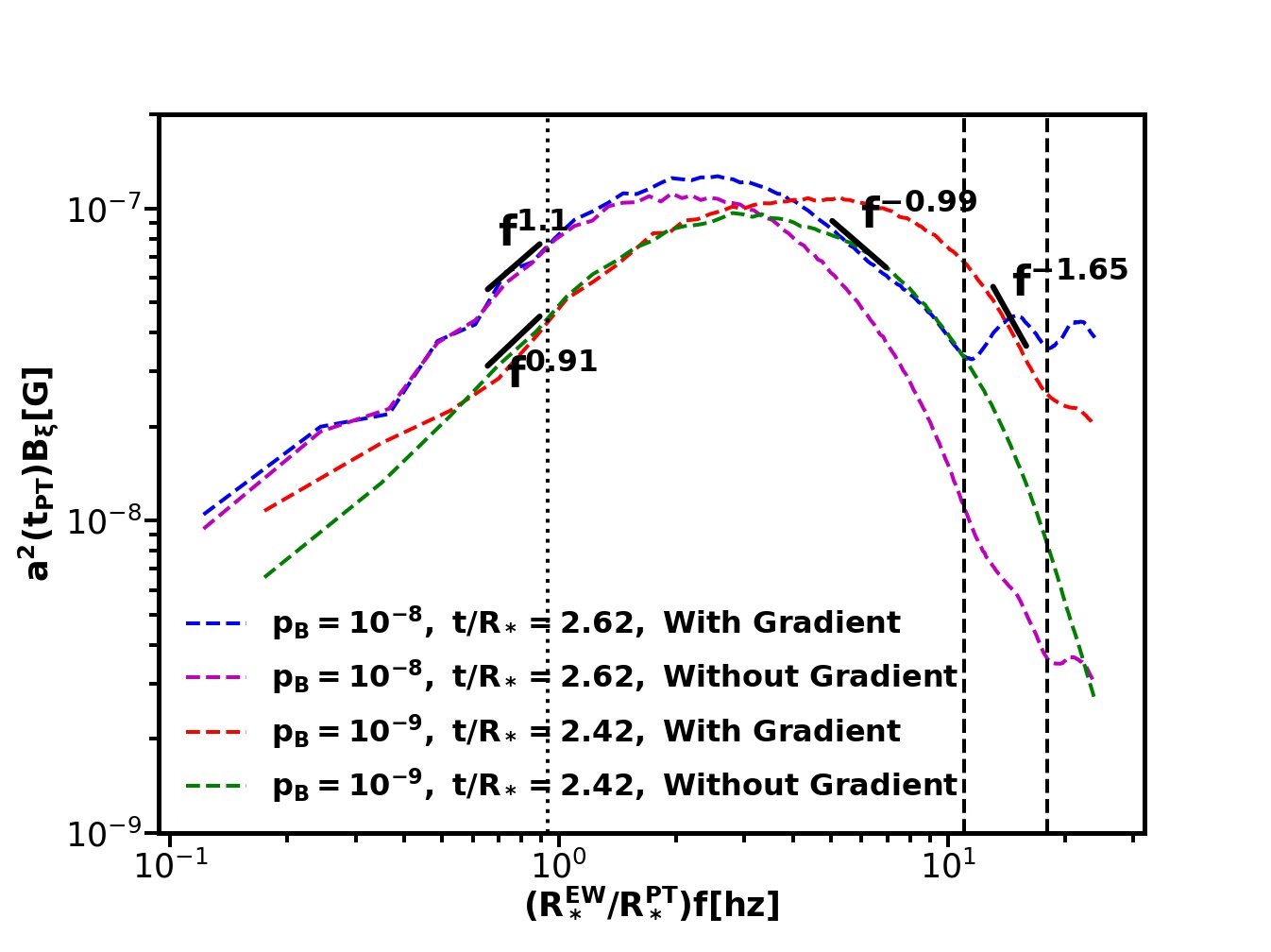}
\includegraphics[width=0.37\textwidth]{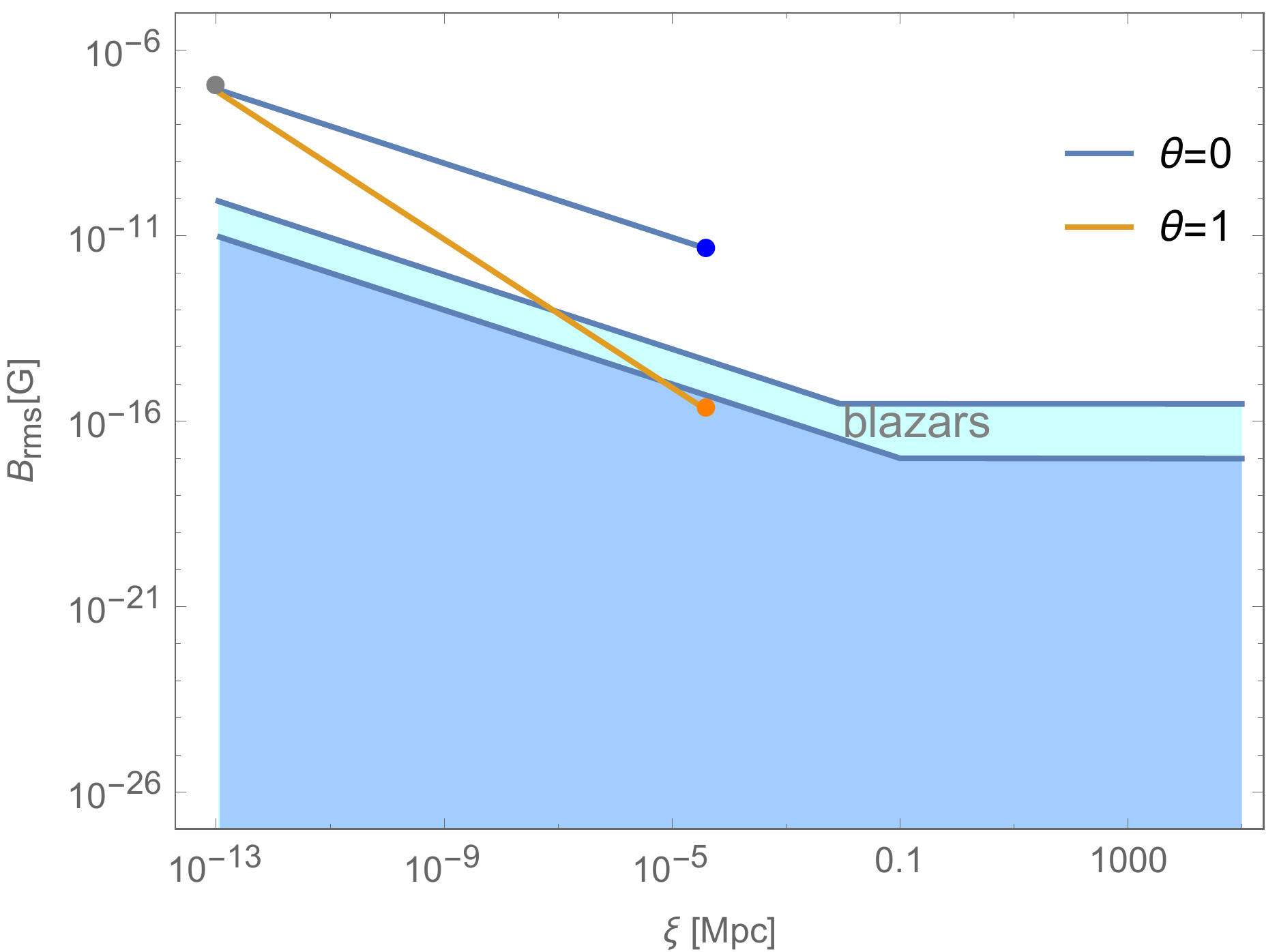}
\caption{Top: the frequency-dependent MF spectrum for different time during the first-order PT with $N_b=160$ (with $\gamma_\star=2.98$) and $N_b=36$ (with $\gamma_\star=4.84$). The length scale associated with the $R_\star$ and $L_w$ are plotted as vertical black dotted line and the vertical dashed line; Bottom: the MF strength as a function of correlation length, with Cyan and Blue region corresponding to lower bounds set by blazars in Ref.\cite{Biteau:2018tmv} and Ref.~\cite{Taylor:2011bn}. We perform the simulation until the time of $t/R_\star=2.42,2.62$ as shown in the top plots, that are long enough for the generation of GWs and MFs, as the stop growth of amplitudes of GW and MF can be found there, see supplemental material. } \label{pmfgw}
\end{center}
\end{figure}

We first numerically calculate the spectrum of the MF strength generated during the first-order electroweak PT when the GWs are produced (we consider the PT temperature $T=100$ GeV). 
As shown in the top panel of Fig.~\ref{pmfgw}, our simulation results in a broken power-law spectrum, with $B_{\xi}\propto f^{0.91(1.1)}$ for $f<f_\star$ (here $f_\star$ indicates the peak frequency) and $B_{\xi}\propto f^{-1.65(-0.99)}$
for $f>f_\star$ for $\gamma_\star=4.84$ with $p_B=10^{-9}$ ( $\gamma_\star=2.98$ with $p_B=10^{-8}$), which means that the spectra of MFs from PT can be used to differentiate different bubble collision scenarios. In comparison with thicker wall ($p_B=10^{-8}$), the MF spectrum falls off more quickly for thin wall scenario ($p_B=10^{-9}$) (See Fig.~2 in the supplemental material for the time evolution of the MF spectra.). The MF strength can reach $B_\xi \sim 10^{-7}$ G at the peak frequency for the thin wall scenario, with the
correlation length being $\xi \sim10^{-7}$ pc. The effects of the Higgs gradients contributions (the terms of the second line in Eq.~\ref{eq:amunu}) to the MF generation are also shown there, which emerges after collisions, increases and dominates the MFs production in the oscillation phase, with the magnitude of MFs increases by around three times after bubble collisions. 
This characteristic is observed for the first time and strongly supports the MFs production mechanism proposed by Vachaspati~\cite{Vachaspati:1991nm,Vachaspati:2001nb}. The MFs spectrum here is different from Ref.~\cite{Zhang:2019vsb} where they didn't consider the effects of the bubble wall thickness. 
Taking into account the effects of the hydromagnetic turbulence for MFs produced in the electroweak PT allow us to constrain the PT 
parameter spaces by the cosmic MFs observations. In bottom panel of Fig.~\ref{pmfgw}, we present the bounds on the MFs from the intergalactic MFs blazar observations for the thin wall scenario, where we adopt scaling law that governs the evolution of MFs and the correlation length for the case where MFs do not have enough time to reach the fully helical stage before recombination, as suggested by the numerical simulation of MFs evolution under the Magnetohydrodynamic turbulence~\cite{Brandenburg:2017neh}: $B_{rms}=B_\star (\xi_M/\xi_\star)^{-(\theta+1)/2}$ for $\theta=0,1$. 
 We didn't present the bounds coming from the Big-Bang Nucleosynthesis~\cite{Kahniashvili:2010wm,Kawasaki:2012va} and the spectrum and anisotropies of the cosmic microwave background measurements~\cite{Seshadri:2009sy,Ade:2015cva,Jedamzik:1999bm,Barrow:1997mj,Durrer:1999bk,Trivedi:2010gi}, since the two yield null result for the current study.

 \begin{figure}[!htp]
\begin{center}
\includegraphics[width=0.4\textwidth]{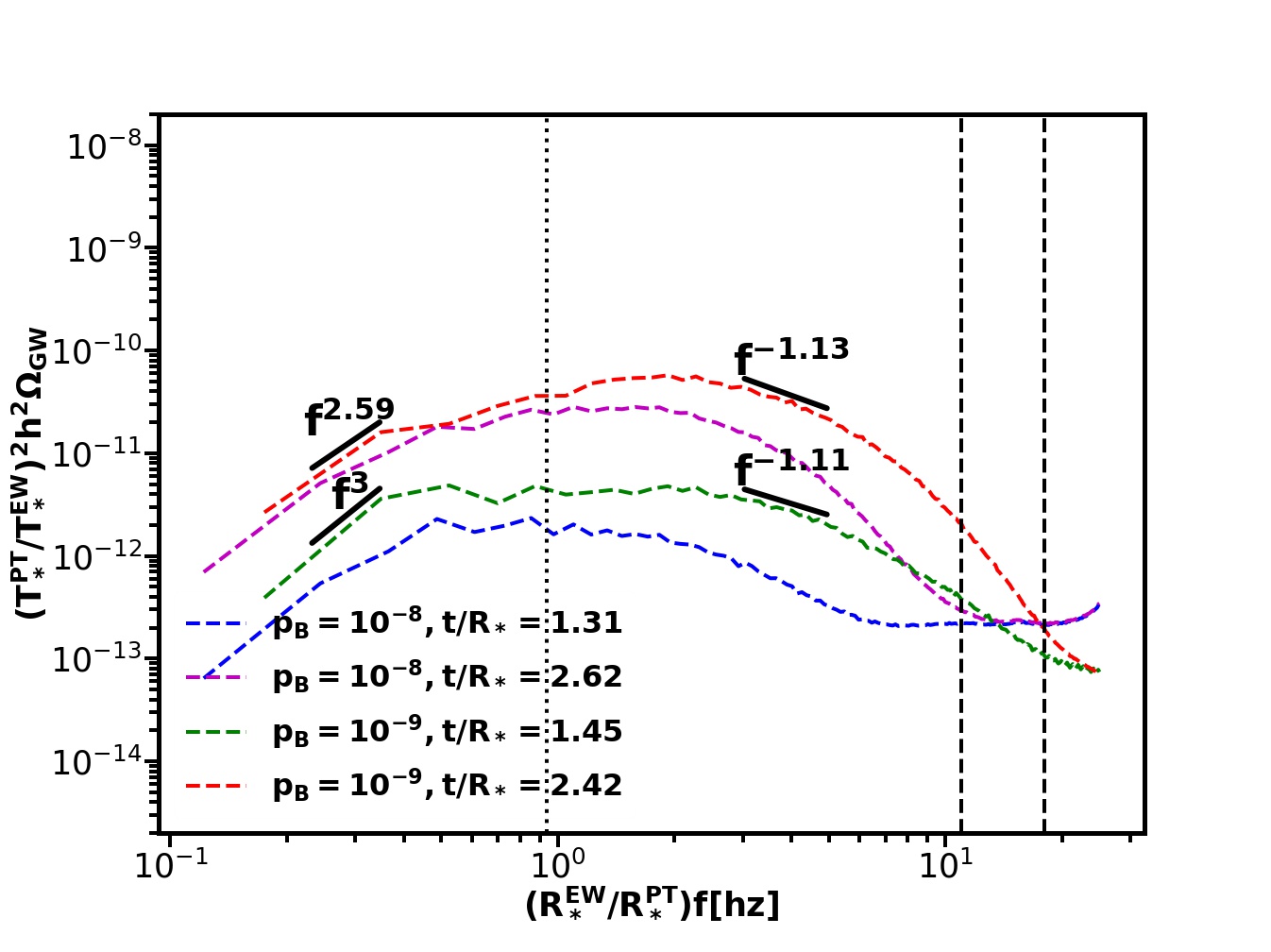}
\includegraphics[width=0.37\textwidth]{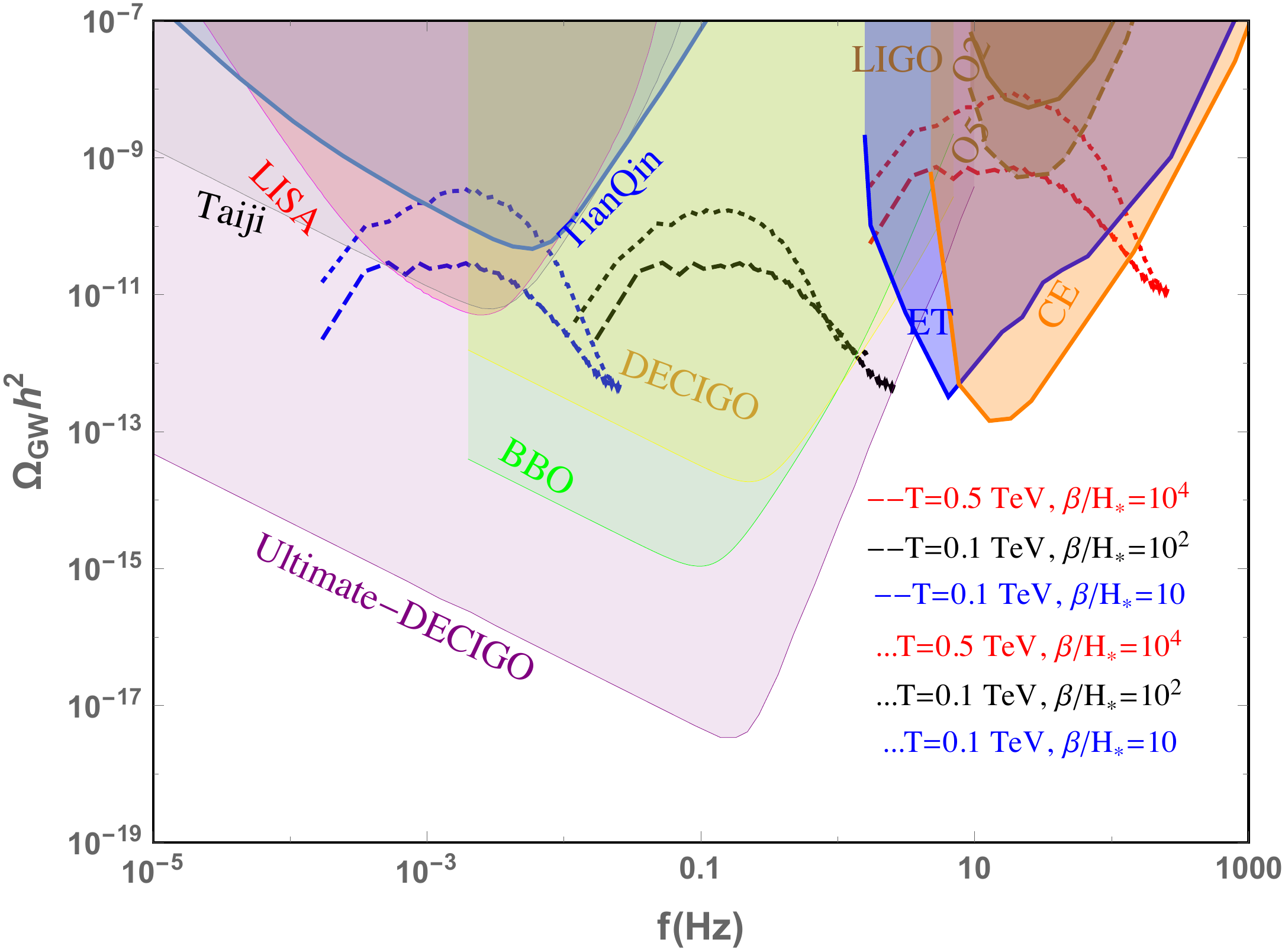}
\caption{Top: The frequency-dependent GW energy spectrum for different time during the first-order PT with $N_b=160$,$\gamma_\star=2.98$, and $N_b=36$,$\gamma_\star=4.84$. The length scale associated with the $R_\star$ and $L_w$ are plotted as vertical black dotted line and the vertical dashed line; Bottom: sensitivity of future GW detectors on the GW spectrum rescaled from the plots in the top panel ($t/R_\star=1.45$ and $t/R_\star=2.42$).  } \label{GWsim}
\end{center}
\end{figure} 

We now turn to study the property of the GW spectrum generated from the first-order electroweak PT. 
To show the wall width effects on the GW spectra, we present the GW production from the bubble collisions during the first-order PT with $p_B=10^{-8,-9}$ in the top panel of Fig.~\ref{GWsim}. In both two cases, the GW energy densities increase by around one order from bubble collision to the oscillation phase where MFs are generated (See Fig.~3 in the supplemental material for the time evolution of the GW spectra). 
Our simulation shows that the MFs contribution to the GWs production is negligible.  
We reconfirm the observation of Ref.~\cite{Child:2012qg,Cutting:2018tjt,Cutting:2020nla}: 1) before $t<R_\star$, the dominant contribution to GWs is from the bubble collisions; 2) after $t>R_\star$, the GW energy spectrum continues growing and the peak of $\Omega_{\mathrm{GW}}$ shifts toward a higher frequency in the oscillation/coalescence phase. We find that the magnitude of the GW spectrum falls into a valley at the length scale of the wall thickness $L_w$, and we do not find the negligible contribution of GWs corresponds to the wall thickness observed in Ref.~\cite{Cutting:2018tjt,Cutting:2020nla}. We find that the GW energy spectrum at $t/R_\star=1.45$ (the bubble collision phase) has a broken power-law form with $\Omega_{\mathrm{GW}}\propto f^{3.0,-1.11}$ for infrared and large frequency regions when $p_B=10^{-9}$, which is consistent with the previous envelope approximation study for the thin wall limit~\cite{Kosowsky:1992vn,Huber:2008hg,Caprini:2009fx,Konstandin:2017sat}, 
where the low-frequency behavior reveals the causality requirement~\cite{Maggiore:2018sht}, and the high-frequency behavior is consistent with the analytical calculations~\cite{Huber:2008hg} in which the single bubble contribution is dominant~\cite{Jinno:2016vai}.
At the simulation time of $t/R_*=2.42$, the GW energy spectrum grows to be $\Omega_{\mathrm{GW}}\propto f^{2.59,-1.13}$ in the oscillation phase. Note that $\Omega_{\mathrm{GW}}\propto f^{3}$ in the far infrared region is a universal result as found in Ref.~\cite{Cai:2019cdl}. Since GWs in the infrared region are sourced at the scale larger than $R_{*}$, in this case $\Omega_{\mathrm{GW}}\propto f^{2.59}$ may result from the limited length scale of the lattice, also may come from the fact that we 
fitted this form in the region not very far from the peak frequency, and we expect $\Omega_{\mathrm{GW}}\propto f^{3}$ could be reproduced with larger resolution simulations and in the lower frequency region. 
Similar to the MF production, the behavior of the high frequency tail of $\Omega_{\mathrm{GW}}$ (top panel of Fig.~\ref{GWsim}) is affected by the oscillations of the Higgs fields in bubbles overlap regions during the coalescence stage after the bubble collision. 
Different PT models can be realized in particular particle physics models beyond the SM, with different PT temperatures ($T$) and PT duration times ($\beta/H$). 
For example, the electroweak PT can be first-order in xSM model and SMEFT, see Ref.~\cite{Zhou:2020idp}, with $T=0.1$ TeV and $\beta/H_*=100$, and the peak frequency of the GW spectrum locates at $f\sim 0.1$Hz. 
To illustrate the sensitivities of variant GW detectors (including LISA~\cite{Audley:2017drz}, Taiji~\cite{Guo:2018npi}, TianQin~\cite{Luo:2015ght}, DECIGO~\cite{Yagi:2011wg}, and BBO~\cite{Corbin:2005ny}, LIGO~\cite{Abbott:2016blz,Thrane:2013oya,LIGOScientific:2019vic}, Einstein Telescope \cite{Hild:2010id, Punturo:2010zz}, and Cosmic Explorer~\cite{Evans:2016mbw}) for the GWs produced from variant PT models, we plot the bottom panel of Fig.~\ref{GWsim}, where the GWs are rescaled from the GW spectra generated with $p_B=10^{-9}$ at $t/R_\star=1.45$ (dashed curves) and $t/R_\star=2.42$ (dotted curves).

\noindent{\it \bfseries Conclusions:}
\label{sec:conclusions}
In this Letter, we performed a simultaneous study of the MFs and GWs production at the first-order electroweak PT. Given the evolution of magnetohydrodynamics turbulence after the PT, we obtain the MFs at the characteristic correlation length~\cite{Kahniashvili:2012uj,Brandenburg:2017neh} that may seed the observed MFs in galaxy clusters. Our simulation suggest that the MFs strength generated from the bubble collisions during the electroweak PT can reach $B_{rms}\sim 10^{-7}$ G at a typical characteristic correlation length $\xi \sim10^{-7}$ pc. The signature can be used to distinguish the MFs from the PT from other sources of MFs in the future observations. We note that though the maximum MFs strength here is comparable with the helical MFs generated by sphaleron decay when CP violation shows up~\cite{Copi:2008he,Vachaspati:2001nb,Cornwall:1997ms}, the MFs spectrum is totally different, which can be probed by future MFs observations. 
Furthermore, the probe of GWs spectrum at future GW detectors (such as LISA) is expected to discriminate GWs from the PT and other GW sources. 
We therefore establish the possibility for probing the first-order electroweak PT directly through the observation MFs at the correlation length, and the detection of its predicted GWs at space-based interferometers.  This  can serve as a inspection of the underlying new physics beyond the Standard model\footnote{For practical models studies of intergalactic MFs and GWs, see Ref.~\cite{Ellis:2019tjf,Ellis:2020uid}.}. 

In the present study, we didn't consider the interaction between fields and plasma since we focus on the MFs and GWs generation during the bubble collision stage. For vacuum phase transition where the sound wave absent, one only needs to consider bubble collisions. Generally, with the interaction between Higgs field and plasma, the predicted GW
signal might be amplified to some extent in the high frequency region in comparison with the case only considering the bubble collision. For that situation, the sound wave will dominate over the GWs from bubble collisions~\cite{Caprini:2019egz}. We expect our observation still holds there and left the detailed study to another publications. Different GW sources are distinguished by their different spectra. For GW spectrum produced by the first-order PT, one of key targets in the space-based GW detectors like LISA~\cite{Audley:2017drz}, it is of typical broken power-law shape with different  slopes for low and high frequencies~\cite{Caprini:2019egz,Caprini:2019pxz,Caprini:2018mtu}.

\noindent{\it \bfseries Acknowledgements}
We thank John T. Giblin, Marek Lewicki, David Weir, Zach Weiner, Yiyang Zhang, and Haipeng An for communications. We are grateful to Francesc Ferrer, and Tanmay Vachaspati for enlightfull duscussions.     
Ligong Bian was supported by the National Natural Science Foundation of China under the grants Nos.12075041, 11605016, and 11947406, and Chongqing Natural Science Foundation (Grants No.cstc2020jcyj-msxmX0814), and the Fundamental Research Funds for the Central Universities of China (No. 2019CDXYWL0029). 
RGC is supported by the National
Natural Science Foundation of China Grants No.11690022, No.11821505, No. 11991052, No.11947302 and by the Strategic
Priority Research Program of the Chinese Academy of Sciences Grant No. XDB23030100 and the Key Research Program
of Frontier Sciences of CAS.

\noindent{\it \bfseries  Supplemental material}
The equations of motions~(EOMs) are given by:
\begin{equation}
\begin{split}
\partial_{0}^{2}\Phi = & D_{i}D_{i}\Phi-\frac{dV(\Phi)}{d\Phi},\\
\partial_{0}^{2}B_{i} = & -\partial_{j}B_{ij}+g'\, \text{Im} [\Phi^{\dagger}D_{i}\Phi ],\\
\partial_{0}^{2}W_{i}^{a} = & -\partial_{k}W_{ik}^{a}-g\, \epsilon^{abc}W_{k}^{b}W_{ik}^{c}
 +g\, \text{Im}[\Phi^{\dagger}\sigma^{a}D_{i}\Phi ]\;,
\label{eq:eom}
\end{split}
\end{equation}
with the solutions subjected to the Gauss constraints:
\begin{equation}
\begin{split}
	\partial_{0}\partial_{j}B_{j}  &- g'\, \text{Im}[\Phi^{\dagger}\partial_{0}\Phi ] = 0,\\
	\partial_{0}\partial_{j}W_{j}^{a} &+g\, \epsilon^{abc}W_{j}^{b}\partial_{0}W_{j}^{c} -
g\, \text{Im}[\Phi^{\dagger}\sigma^{a}\partial_{0}\Phi ] = 0.
\label{eq:gauss-su2}
\end{split}
\end{equation}
The potential energy is given by $V(\phi)$, which we take the SM potential extended by a dimensional six operator $(\Phi^{\dagger}\Phi)^3/\Lambda^2$ for the evolution of these EOMs, the case of $\Lambda=600$ GeV  can account for the first-order EWPT~\cite{Zhou:2019uzq}. The gradient energy and kinetic energy are $\rho_{\rm Gradient}=(D_0 \Phi)^\dagger(D_0 \Phi)$ and $\rho_{\rm Kinetic}=(D_i \Phi)^\dagger(D_i \Phi)$ ($i=1,2,3$) with the $D$ being the covariant derivative, the two together construct the kinetic term of the electroweak theory. 

\begin{figure}[!htp]
\begin{center}
\includegraphics[width=0.2\textwidth]{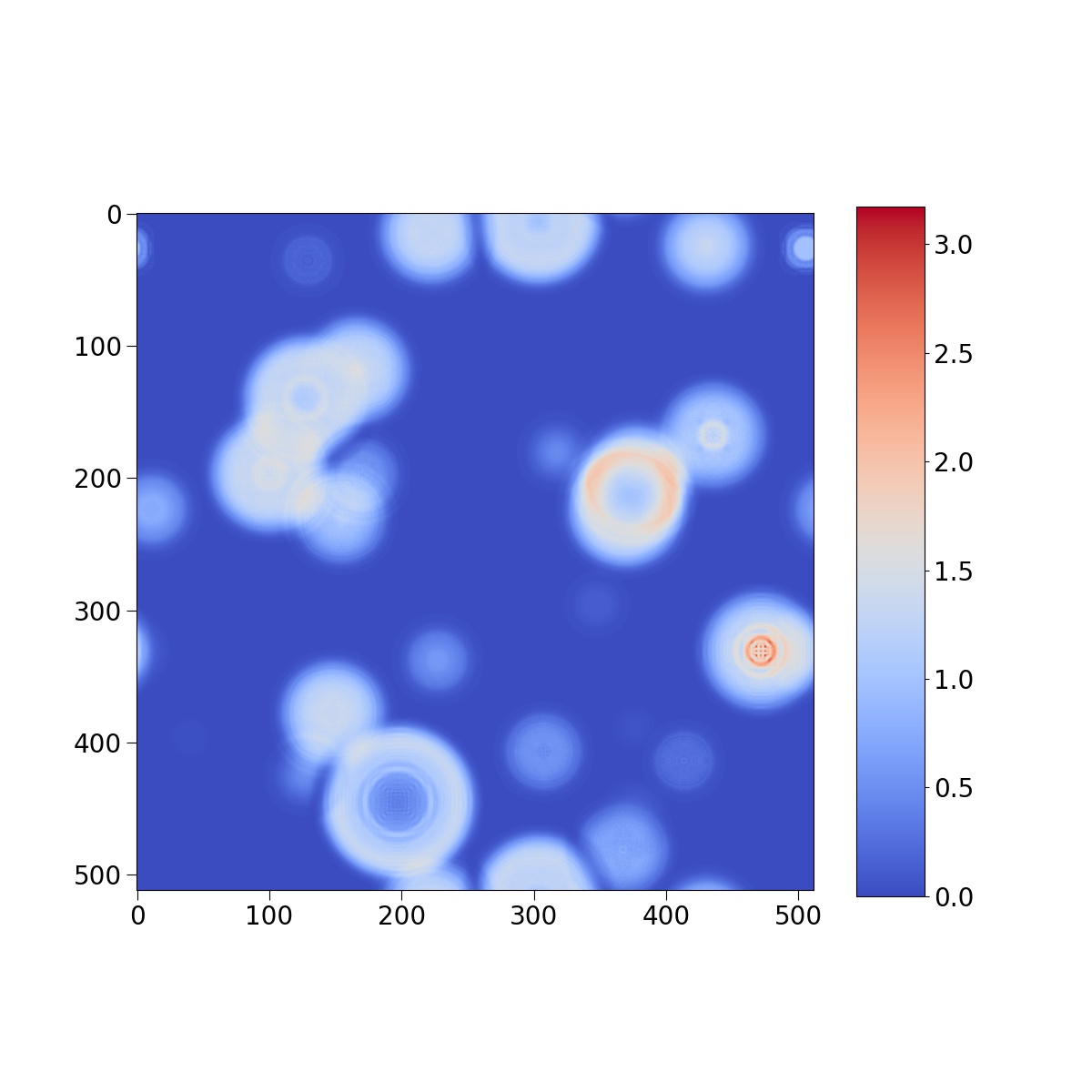}
\includegraphics[width=0.2\textwidth]{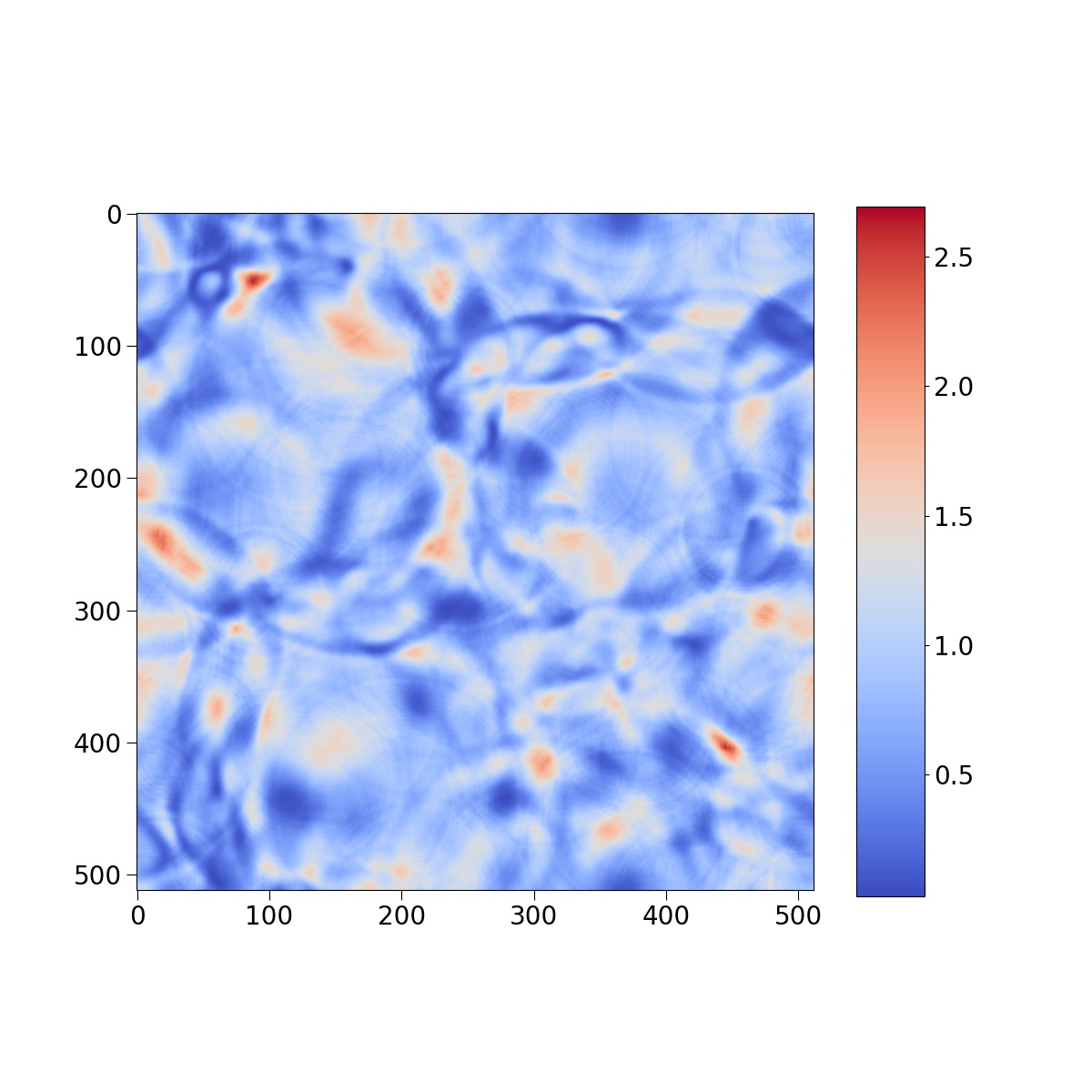}
\includegraphics[width=0.2\textwidth]{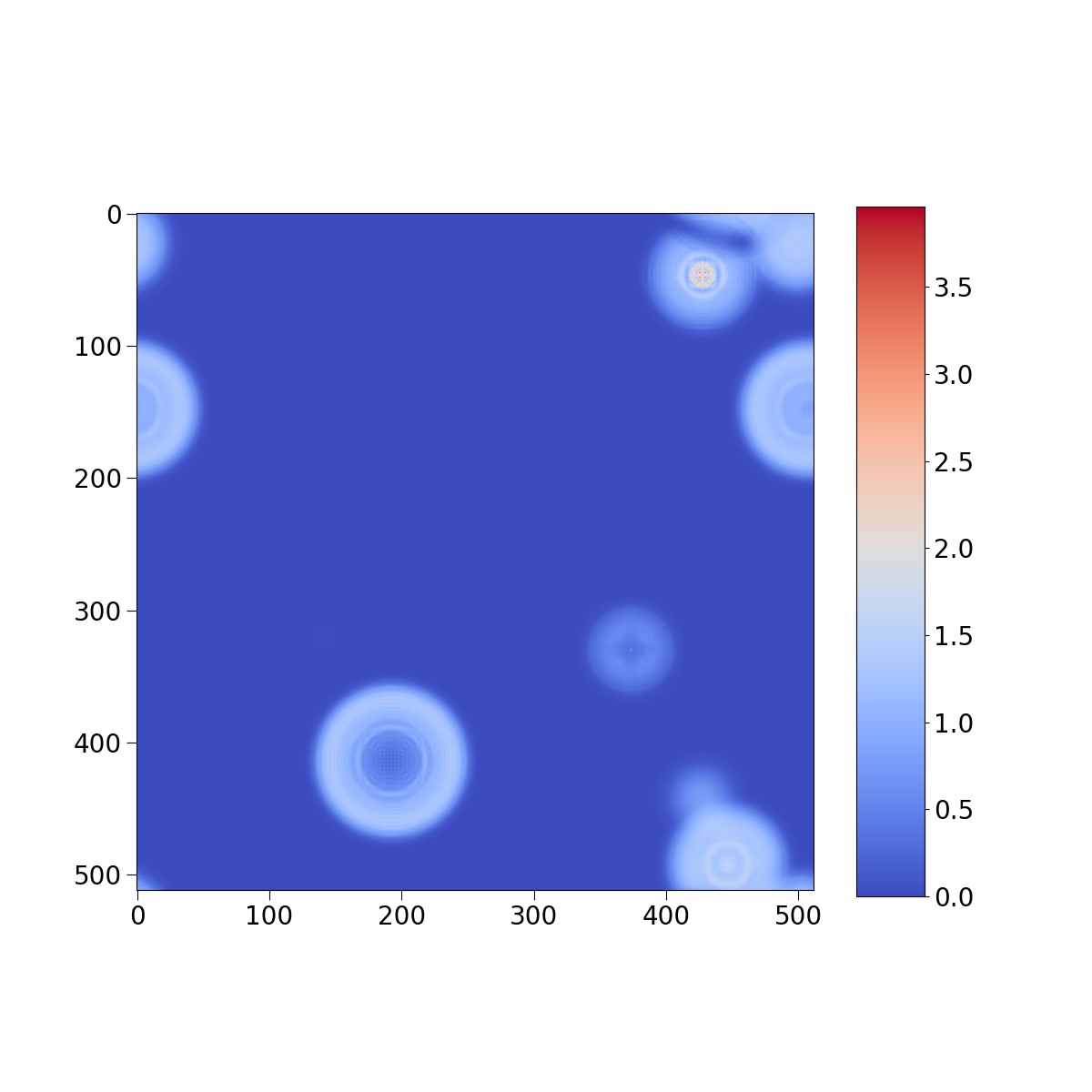}
\includegraphics[width=0.2\textwidth]{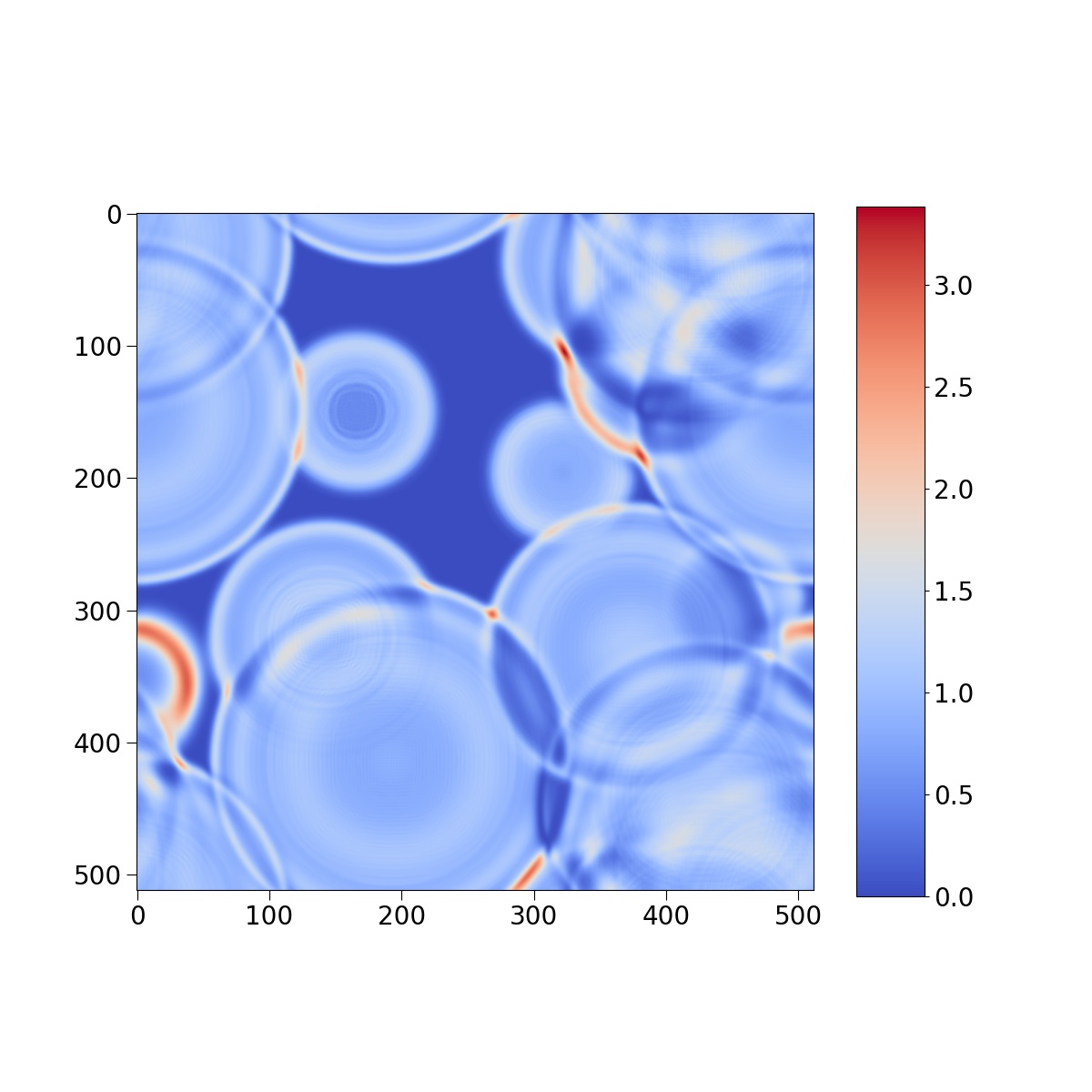}
\caption{The distributions of $|\Phi|^2/v^2$ at $200$ and $500$ steps for bubble nucleation rates $p_B=10^{-8}$ (top panel) and $p_B=10^{-9}$ (bottom panel). } \label{fig:bubble}
\end{center}
\end{figure}

Fig.~\ref{fig:bubble} shows the expansion and collision
of the randomly generated bubbles, where we present the cases with bubble nucleation rates $p_B=10^{-8,-9}$. The relation among the bubble velocity, mean bubble separation,  and phase transition duration (that can be obtained through a bounce solution of the Euclidean action for  nucleating a critical bubble of broken phase at finite temperature~\cite{Caprini:2019egz})\footnote{The relations among these parameters in particular PT models simulations can be obtained as in Ref~\cite{Hindmarsh:2019phv,Cutting:2020nla,Cutting:2018tjt,Hindmarsh:2013xza} . } .

\begin{figure*}[!htp]
\begin{center}
\includegraphics[width=0.4\textwidth]{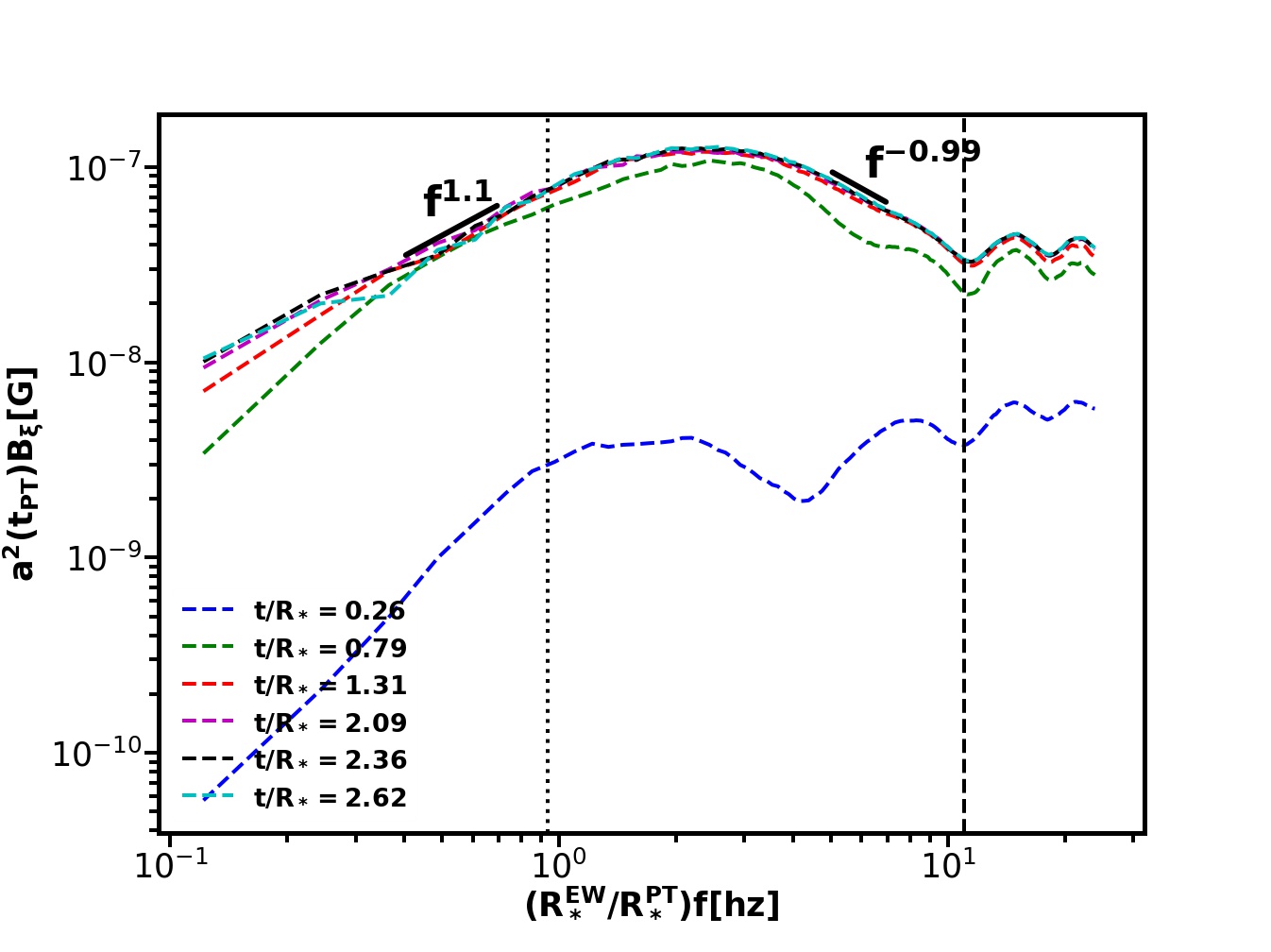}
\includegraphics[width=0.4\textwidth]{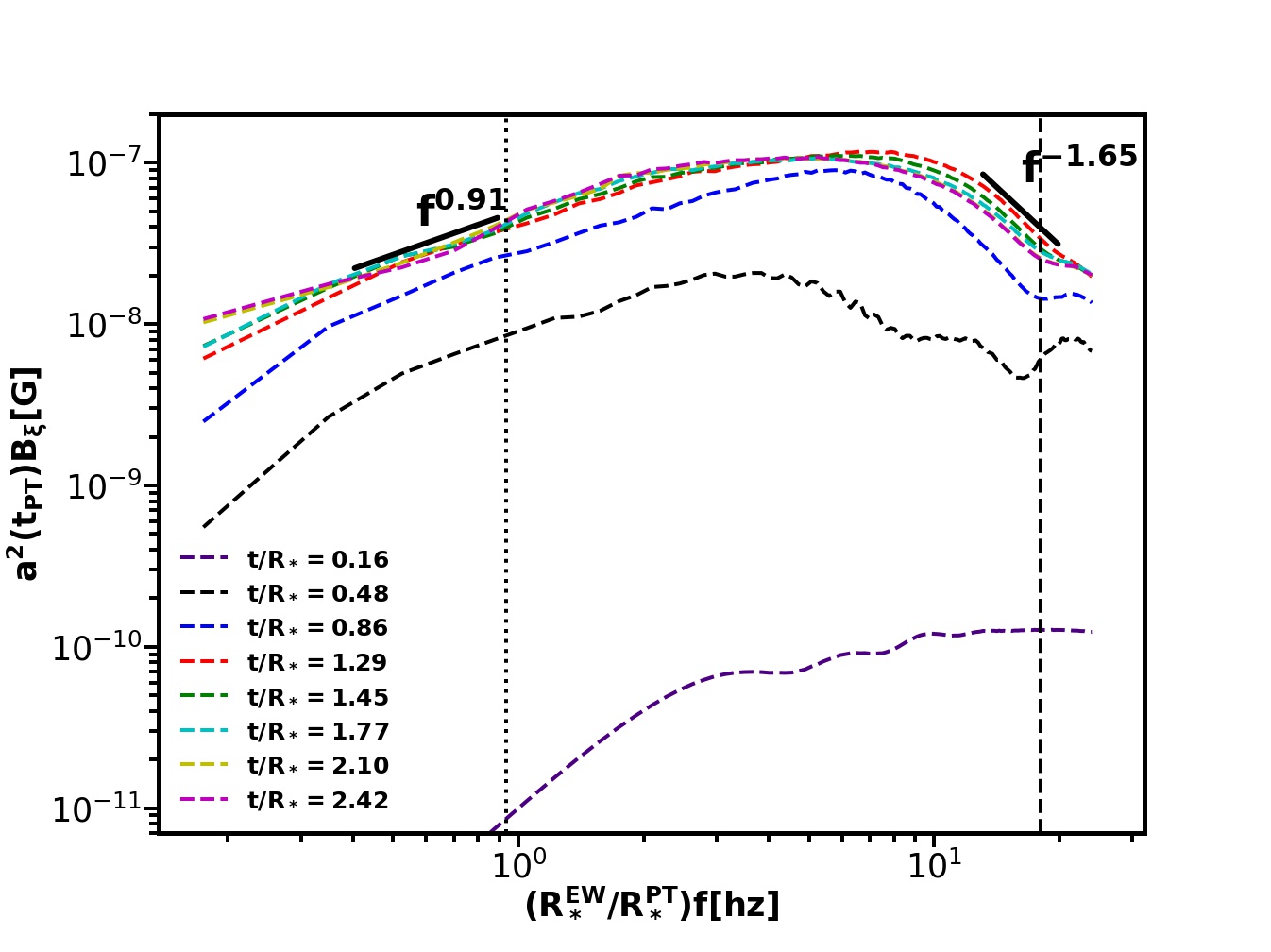}
\caption{The frequency-dependent MF spectrum for different time during the first-order PT with $N_b=160$(with $\gamma_\star=2.98$) and $N_b=36$ (with $\gamma_\star=4.84$) in left and right panels. The length scales associated with the $R_\star$ and $L_w$ are plotted as vertical black dotted line and the vertical dashed line. } \label{pmf}
\end{center}
\end{figure*}

\begin{figure*}[!htp]
\begin{center}
\includegraphics[width=0.4\textwidth]{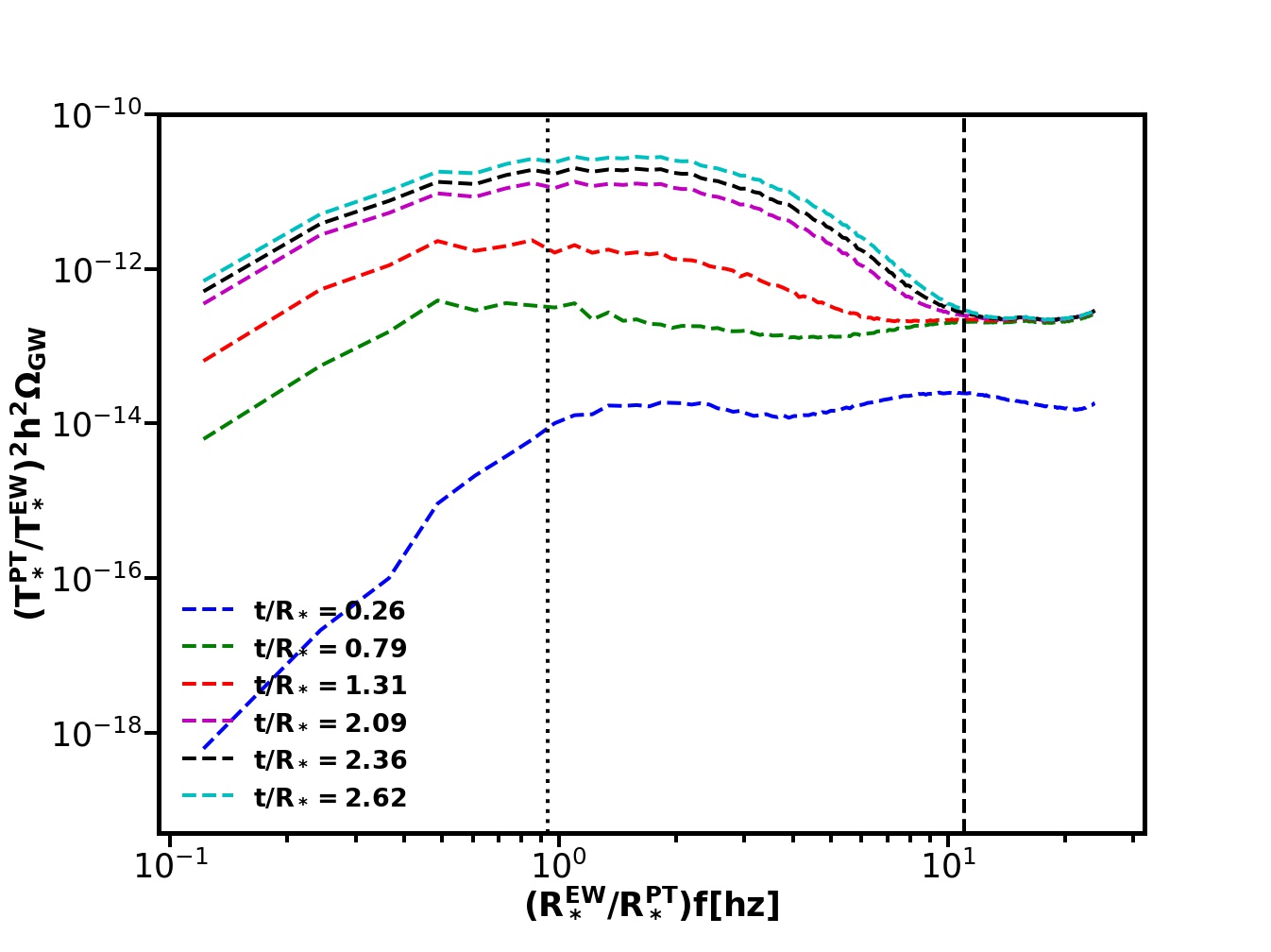}
\includegraphics[width=0.4\textwidth]{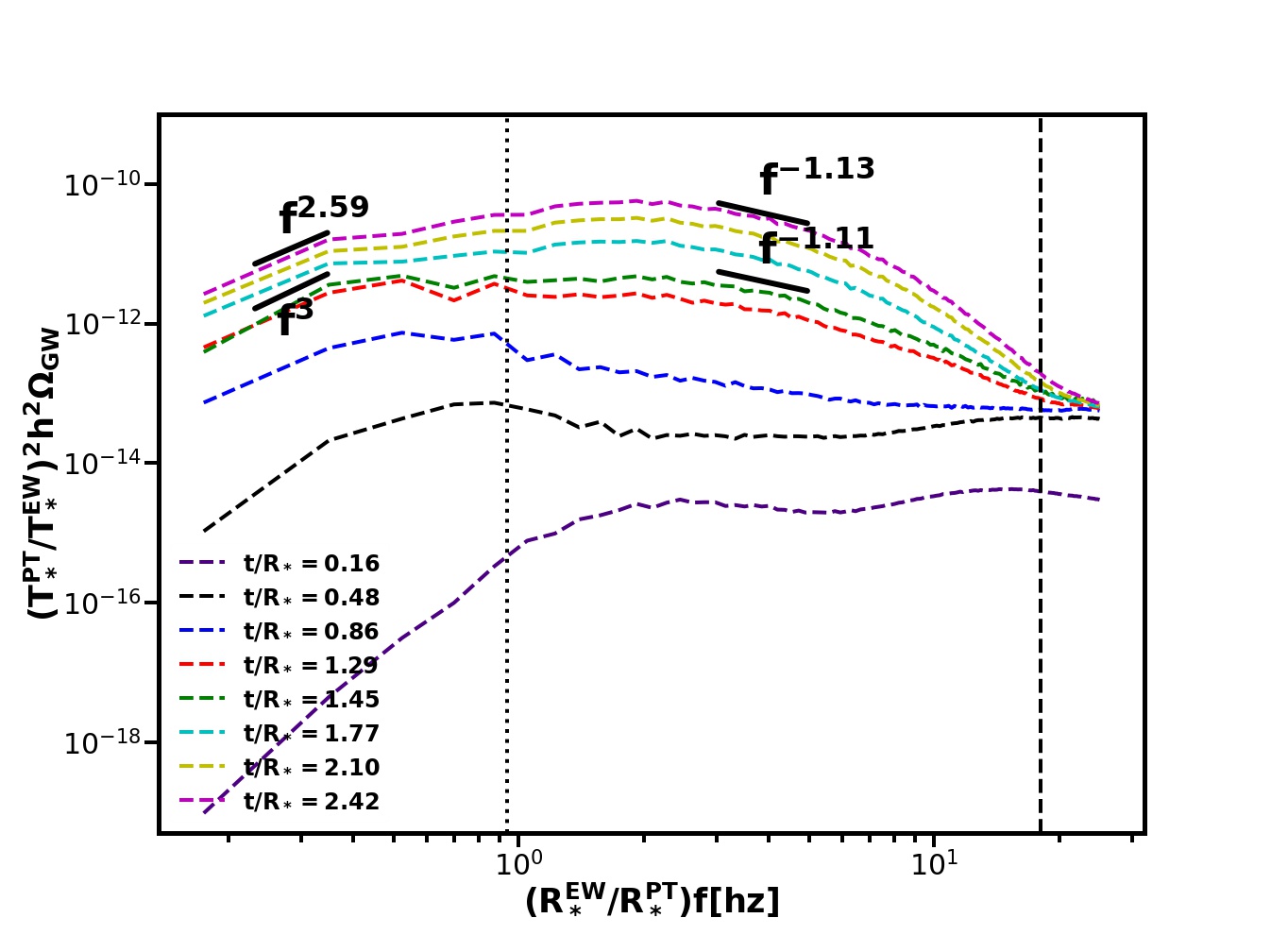}
\caption{Left: The frequency-dependent GW energy spectrum for different time during the first-order PT with $N_b=160$,$\gamma_\star=2.98$; Right: The frequency-dependent GW energy spectrum for different time during the first-order PT with $N_b=36$,$\gamma_\star=4.84$. The length scales associated with the $R_\star$ and $L_w$ are plotted as vertical black dotted line and the vertical dashed line.  } \label{GWsim2}
\end{center}
\end{figure*}

In Fig.~\ref{pmf}, we plot the time accumulation MF spectra for $p_B=10^{-8}$ and $p_B=10^{-9}$ scenarios to consider thick and thin wall effects on the MF production. The MF spectra grow slightly during the Higgs oscillation stage.
In  Fig.~\ref{GWsim2}, we plot the time accumulation GW spectra for $p_B=10^{-8}$ and $p_B=10^{-9}$ scenarios to consider thick and thin wall effects on the GWs production.
We remove simulation artifacts by cut the spectra at around $R_\star^{EW}/R_\star^{PT} f[hz]\sim$ 20. In the right panel, we can find that the amplitude of GWs continually grows and the GW spectra change during the coalescence phase due to the Higgs oscillation in bubble overlap regions. Specially, the scaling of the high frequency tails decreases slightly. 

\bibliographystyle{apsrev}

\bibliography{GWBPT}

\end{document}